# OBSERVED MOBILITY BEHAVIOR DATA REVEAL "SOCIAL DISTANCING INERTIA"

Sepehr Ghader, Research Scientist
Maryland Transportation Institute, Department of Civil and Environmental Engineering
1173 Glenn Martin Hall, University of Maryland
College Park, MD 20742, Email: sghader@umd.edu

Jun Zhao, Graduate Research Assistant
Maryland Transportation Institute, Department of Civil and Environmental Engineering
1173 Glenn Martin Hall, University of Maryland
College Park, MD 20742, Email: jzhao124@umd.edu

Minha Lee, Graduate Research Assistant
Maryland Transportation Institute, Department of Civil and Environmental Engineering
1173 Glenn Martin Hall, University of Maryland
College Park, MD 20742, Email: minhalee@umd.edu

Weiyi Zhou, Graduate Research Assistant
Maryland Transportation Institute, Department of Civil and Environmental Engineering
1173 Glenn Martin Hall, University of Maryland
College Park, MD 20742, E-mail: wyzhou93@umd.edu

Guangchen Zhao, Graduate Research Assistant
Maryland Transportation Institute, Department of Civil and Environmental Engineering
1173 Glenn Martin Hall, University of Maryland
College Park, MD 20742, Email: gczhao@umd.edu

Lei Zhang, Herbert Rabin Distinguished Professor (Corresponding Author)
Director, Maryland Transportation Institute
Department of Civil and Environmental Engineering
1173 Glenn Martin Hall, University of Maryland
College Park, MD 20742, Email: lei@umd.edu

**ABSTRACT**

The research team has utilized an integrated dataset, consisting of anonymized location data, COVID-19 case data, and census population information, to study the impact of COVID-19 on human mobility. The study revealed that statistics related to social distancing, namely trip rate, miles traveled per person, and percentage of population staying at home have all showed an unexpected trend, which we named "social distancing inertia". The trends showed that as soon as COVID-19 cases were observed, the statistics started improving, regardless of government actions. This suggests that a portion of population who could and were willing to practice social distancing voluntarily and naturally reacted to the emergence of COVID-19 cases. However, after about two weeks, the statistics saturated and stopped improving, despite the continuous rise in COVID-19 cases. The study suggests that there is a natural behavior inertia toward social distancing, which puts a limit on the extent of improvement in the social-distancing-related statistics. The national data showed that the inertia phenomenon is universal, happening in all the U.S. states and for all the studied statistics. The U.S. states showed a synchronized trend, regardless of the timeline of their statewide COVID-19 case spreads or government orders.

# 1. INTRODUCTION AND LITERATURE REVIEW

COVID-19 pandemic caused various societies to deviate from their normal. Many societies faced an unprecedented situation, for which they were not prepared. This event was the first to cause such a large-scale halt on people's normal behavior. Decision makers were not fully aware about people's reaction to new stay-at-home or social distancing orders, as there was no sufficient previous evidence. Now, many weeks into the pandemic, we have learned a lot, not just about the virus, but also about societies' reaction to such events. We need to learn from this experience, so that we can be better prepared as individuals, communities, or societies, the next time such a disastrous event happens. There are a lot to learn on the subject of human behavior amid a pandemic. In this paper, we focus on American's reaction to social distancing. For the first time in history, large scale national evidence is available on this subject. Reaction to social distancing can be studied in-depth through detailed surveys. However, such surveys are costly, and time-consuming to design, collect, and analyze. Furthermore, respondents may have reservations against reporting discord with social distancing orders. Thanks to the technology, a massive amount of data has been passively collected from mobile devices, which can provide observed evidence on how people reacted to social distancing. We used daily feeds of mobile device location data, representing movements of more than 100 Million anonymized devices, integrated with COVID-19 case data from John Hopkins University and census population data to monitor the mobility trends in United States and study social distancing behavior [1].

Mobile device location data are becoming popular for studying human behavior, specially mobility behavior. Earlier studies with mobile device location data were mainly using GPS technology, which is capable of recording accurate information including, location, time, speed, and possibly a measure of data quality [2]. Later, mobile phones and smartphones gained popularity, as they could enable researchers to sudy individual-level mobility patterns [3-5]. Other emerging mobile device location data sources such as call detail record (CDR) [6-8], Cellular network data [9], and social media location-based services [10-14] have also been used by the researchers to study mobility behavior. Mobile device location data has proved to be a great asset for decision-makers amid the current COVID-19 pandemic. Many companies such as Google, Apple, or Cuebiq have already utilized location data to produce valuable information about mobility and economic trends [15-17]. Researchers have also utilized mobile device location data for studying COVID-19- related behavior [18,19]. Our paper utilizes mobile device location data to study social distancing behavior. Non-pharmaceutical interventions such as social distancing are important and effective tools for preventing virus spread. Researchers have highlighted the importance of social distancing in disease prevention through modeling and simulation [20-23]. The simulation models assume a level of compliance, which can now be validated through observed data. One of the most recent studies projected that the recurrent outbreaks might be observed this winter based on pharmaceutical estimates on COVID-19 and other coronaviruses, so prolonged or intermittent social distancing may be required until 2022 without any interventions [24], highlighting the importance of improving our understanding about individual's reaction to social distancing.

In order to study the impact of COVID-19, we have processed a set of national mobile device location data and created an online platform [25]. The next section describes the platform. **Section 3** briefly describes the methodology for processing location data and producing the statistics of interest. The analysis of the results from the platform showed an interesting trend, which we named "social distancing inertia". **Section 4** introduces this phenomenon. **Section 5** discusses the universality of the observed phenomena. The last section is dedicated to discussion and conclusion.

## 2. UNIVERSITY OF MARYLAND COVID-19 IMPACT ANALYSIS PLATFORM

The COVID-19 Impact Analysis Platform, available at [data.covid.umd.edu](data.covid.umd.edu) provides comprehensive data and insights on COVID-19's impact on mobility, economy, and society with daily data updates. Researchers at the University of Maryland (UMD) are exploring how social distancing and stay-at-home orders are affecting travel behavior, spread of the coronavirus, and local economies. We are also studying the multifaceted impact of COVID-19 on our lives, health, economy, and society. Through this interactive analytics platform, we are making our data and research findings available to other researchers, agencies, non-profits, media, and the general public. The platform will evolve and expand over time as new data and impact metrics are computed and additional visualizations are developed. **Table 1** shows the current metrics available in the platform at the national, state, and county levels in the United States with daily updates.

**Table 1. List of metrics available on the COVID-19 impact analysis platform**

| Current Metrics | Description |
|---|---|
| social distancing index | An integer from 0~100 that represents the extent residents and visitors are practicing social distancing. "0" indicates no social distancing is observed in the community, while "100" indicates all residents are staying at home and no visitors are entering the county. |
| % staying home | Percentage of residents staying at home |
| #trips/person | Average number of trips taken per person. |
| % out-of-county trips | The percent of all trips taken that travel out of a county. Additional information on the origins and destinations of these trips at the county-to-county level is available, but not currently shown on the platform. |
| miles traveled/person | Average person-miles traveled on all modes (car, train, bus, plane, bike, walk, etc.) |
| #work trips/person | Number of work trips per person (where a "work trip" is defined as going to or coming home from work) |
| #non-work trips/person | Number of non-work trips per person. (e.g. grocery, restaurant, park, etc.). Additional information on trip purpose (restaurant, shops, etc.) is available, but not currently shown on the platform. |
| #COVID-19 cases | Number of new confirmed COVID-19 cases from the Johns Hopkins University's GitHub repository. |
| population | Number of residents in a nation, state, or county as reported from the national Census database. |

## 3. METHODOLOGY

The research team first integrated and cleaned location data from multiple sources representing person and vehicle movements in order to improve the quality of our mobile device location data panel. We then clustered the location points into activity locations and identified home and work locations at the census block group (CBG) level to protect privacy. We examined both temporal and spatial features for the entire activity location list to identify home CBGs and work CBGs for workers with a fixed work location. Next, we applied previously developed and validated algorithms [26] to identify all trips from the cleaned data panel, including trip origin, destination, departure time, and arrival time. If an anonymized individual in the sample did not make any trip longer than one-mile in distance, this anonymized individual was considered as staying at home. A multi-level weighting procedure expanded the sample to the entire population, using device-level and trip-level weights, so the results are representative of the entire population in a nation, state, or county. The data sources and computational algorithms have been validated based on a variety of independent datasets such as the National Household Travel Survey and American Community Survey, and peer reviewed by an external expert panel in a U.S. Department of Transportation Federal Highway Administration's Exploratory Advanced Research Program project, titled "Data analytics and modeling methods for tracking and predicting origin-destination travel trends based on mobile device data" [26]. Mobility metrics were then integrated with COVID-19 case data, population data, and other data sources. **Figure 1** shows a summary of the methodology. Additional details can be found in a separate paper by the authors.

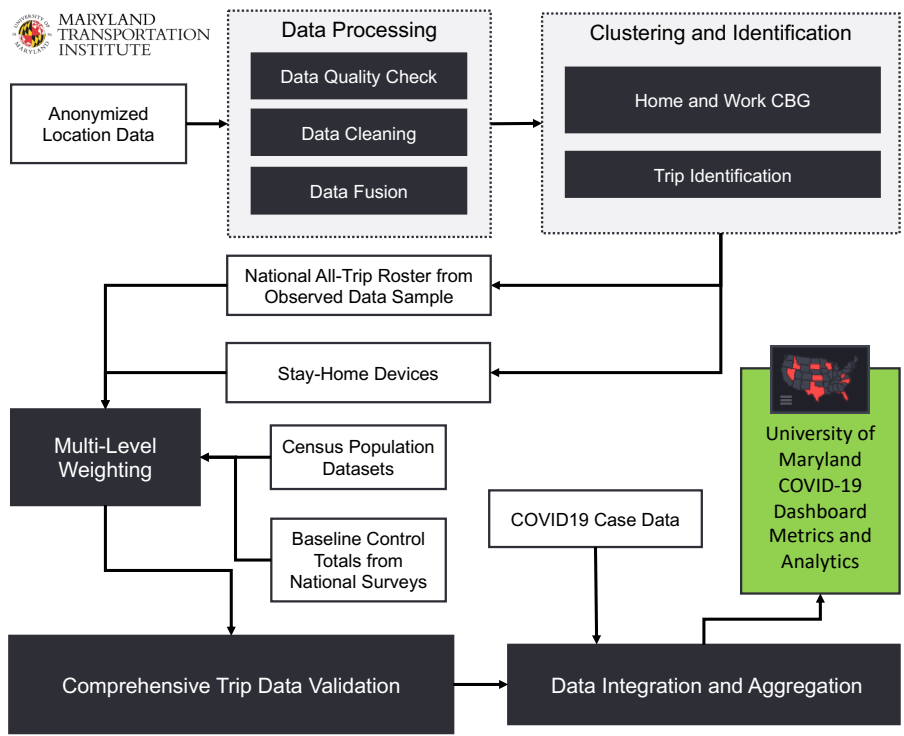

**Figure 1. Methodology**

4. **THE PHENOMENON OF SOCIAL DISTANCING INERTIA**

The analysis of data from the COVID-19 impact analysis platform showed a notable trend. We could observe that as soon as COVID-19 cases first began to appear in significant numbers (i.e., early to mid-March), measures related to social distancing (index, %staying home, #trips per person, trip distance, out-of-town trips, etc.) began to improve quickly, with or without government social distancing orders, suggesting that those who wanted and were able to limit their interactions with others quickly and naturally responded to the emergence of new cases and adopted social distancing practices. However, after about two weeks into the pandemic (mid to late-March), measures related to social distancing stopped improving despite continuously increasing COVID-19 case numbers and government stay-at-home orders. The trends showed that all measures related to social distancing saturated and stopped improving, revealing a phenomenon we name "Social Distancing Inertia." For instance, as observed in **Figure 2** the percentage of people staying home nationwide rapidly increased from 20% to 35% at the onset of COVID-19 and then has stagnated at 35% for three weeks as of April 10. Digging deeper, we observed that the same trend is also observable in the states with the highest number of cumulative cases (**Figure 2**). Here, we focus on three selected statistics of interest related to social distancing. The first is percentage of people staying home, which shows the proportion of population that did not make any trips longer than one mile on a given day (**Figure 2**). The second is number of trips per person, which shows on average how many trips are observed per person on a given day (**Figure 3**). The third is miles traveled per person, which shows on average how many miles is taken per person on a given day (**Figure 4**). In the three figures below, the red curve shows the normalized trend of new confirmed COVID-19 cases, based on the John Hopkin's COVID-19 database [1]. The other curve in each graph shows one of the statistics of interest. The left axis shows the values for the statistics of interest. We have removed weekends to have a consistent comparison. We have also used three day moving averages to remove the day-to-day noises. We can see that in all three figures, in both national and state level results, the static of interest has started increasing around the same time and reached a plateau after about two weeks.

It is also worthwhile to see the effect of government stay-at-home orders. The date of government stay-at-home orders are identified with a black line in the figures. No significant improvement can be observed in any of the statistics after government orders. In fact, the trends have stopped increasing after the orders. Government advisories and stay-at-home orders have not accomplished expected changes in mobility behavior according to our data analysis. The analysis suggests that those who were able to adopt social distancing practices had already done so before government intervention. Those who could not or did not want to stay home showed significant behavior inertia and rendered government stay-at-home orders less effective than expected. The average increase of percentage of people staying at home one week after the order compared to one week before the order is 4.3 percentage points.

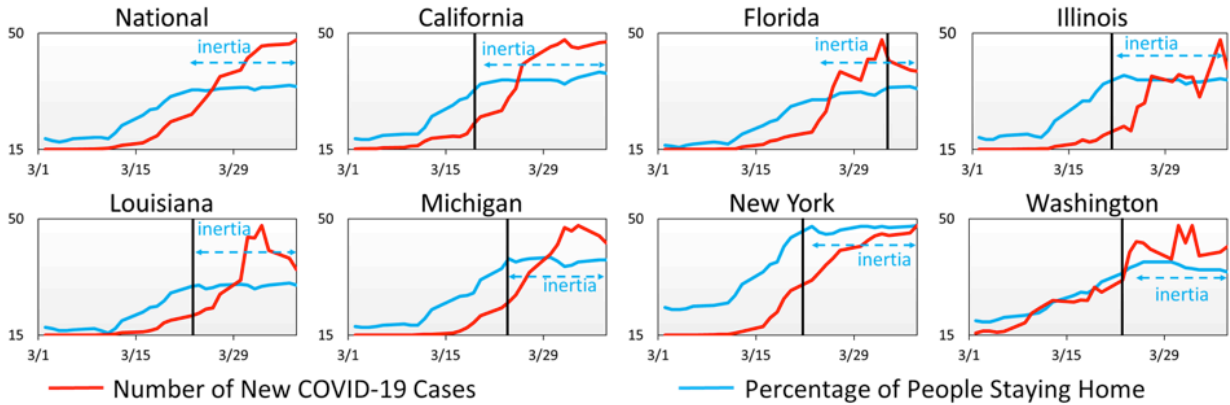
Figure 2. Inertia for percentage of people staying home

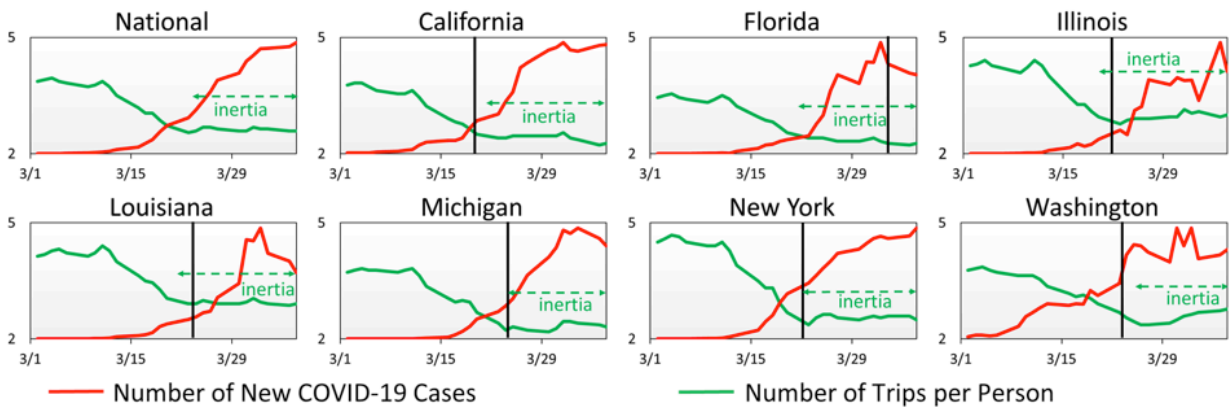
Figure 3. Inertia for number of trips per person

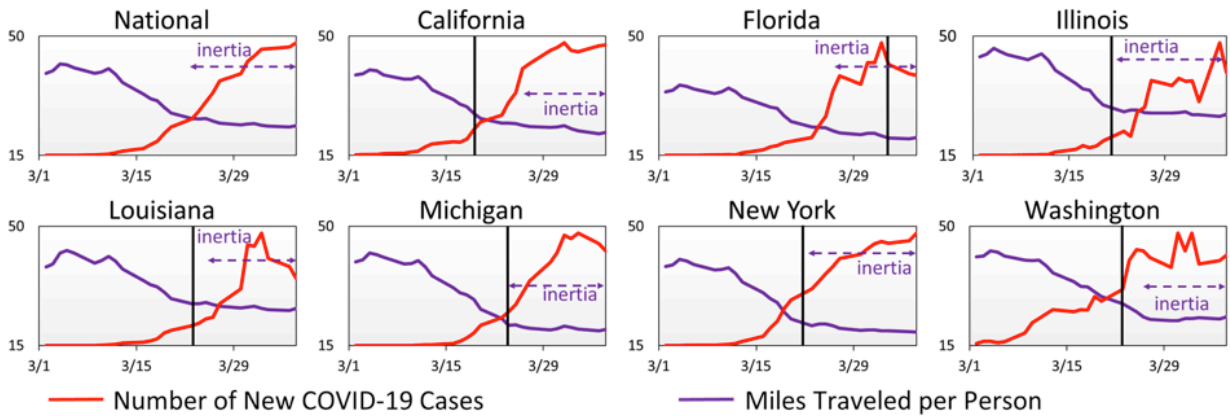
Figure 4. Inertia for miles traveled per person

## 5. THE UNIVERSALITY OF SOCIAL DISTANCING INERTIA

Another interesting observation was the universality of social distancing inertia. Despite the fact that COVID-19 case emergence had a different timeline in each state and government advisory orders were issued in different dates, all states were synchronized in their trends. **Figure 5** shows the percentage of people staying at home. The curves for all states are plotted together to show the universality of the phenomenon, and highlight how synchronized the curves are. **Figure 6** and **Figure 7** are plotted similarly for number of trips per person and miles traveled per person respectively. In these figures, the scales vary between the states, as the mobility behavior is generally different between states that have more urban regions and those that are mainly suburban and rural, but the trends are similar. The figures suggest that the phenomenon is nationwide. The synchrony suggests that people reacted to national emergence of cases or the national emergency declaration on March 13th, not the trend of cases in their own region.

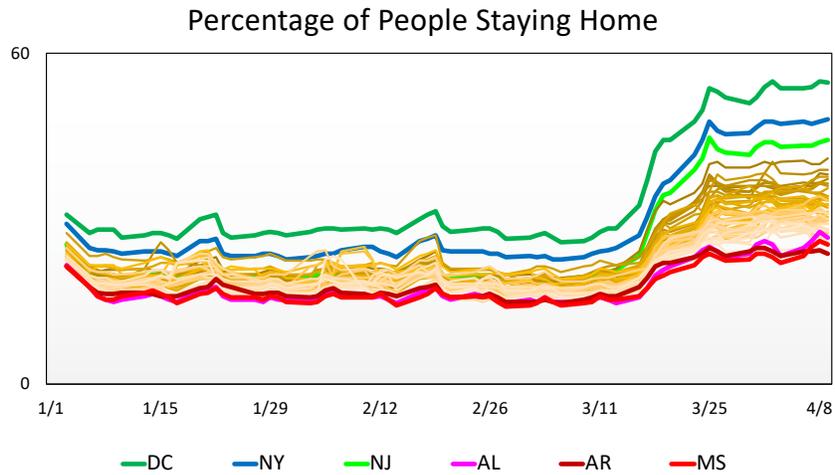

**Figure 5. Universality of inertia for percentage of people staying home**

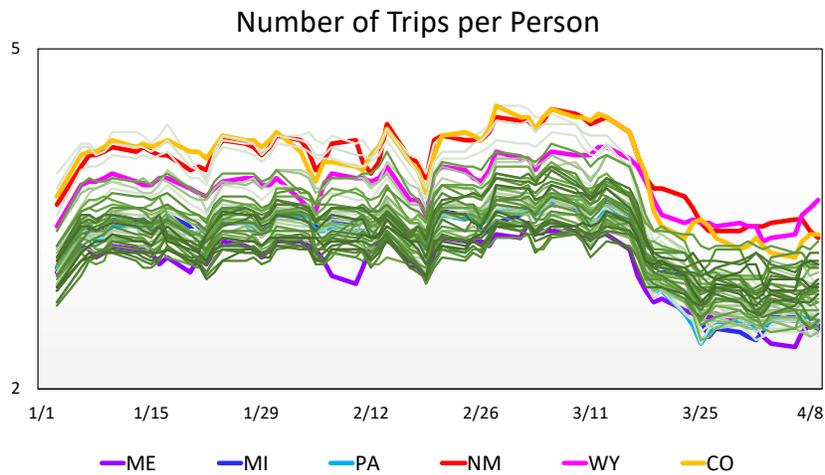

**Figure 6. Universality of inertia for number of trips per person**

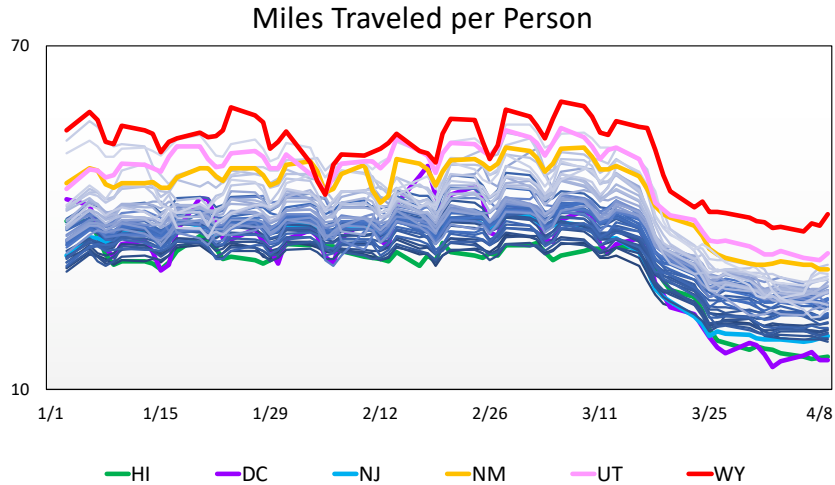

**Figure 7. Universality of inertia for miles traveled per person**

## 6. DISCUSSION

The results showed that the improvements in statistics related to social distancing were smaller than expected. They also showed that the trends reached a plateau after about two weeks. Even though all aspects of our results are suggesting the inertia, they cannot prove that people are not conforming with social distancing orders. One can still argue that the remaining trips are essential trips; or people that are not staying home are essential workers or are people traveling for their essential needs. However, digging deeper can provide further evidence suggesting inertia. The drop in the number of work trips per person was about 50%, which shows that even amid the pandemic, about 50% of people are going to their routine work place. The drop in number of non-work trip was limited to 22%, suggesting that people have not significantly decreased their non-work trip rate. One may argue that the nature of the non-work trips happening amid pandemic may be different from the non-work trips happening before, and the trips amid pandemic may be short walks in the neighborhood. However, looking into the drop in average miles traveled per person, the national drop is limited to 40%. We can see that people are still traveling significant miles. A recent analysis of PlaceIQ data during pandemic showed that foot traffic to groceries has only decreased 27% [27]. Altogether, the result suggest that a significant part of the nation is showing inertia toward social distancing.

Social distancing is very important to slow down the spread of COVID-19. However, performing social distancing is quite different from peoples' habitual behavior. Social distancing policy clashes with the deep-seated human instinct to connect with others in order to regulate emotions, cope with stress, and remain resilient during difficult times [28,29]. Therefore, without enforcement, incentives, or proper knowledge about the risks involved, people have an inertia to keep their habitual behavior. Even though staying home may help prevent disease spread, people tend to repeat previous behaviors [30]. The inertia is common in other aspects of human behavior. In stock market, investors who missed an opportunity to leave a 'bear market' are less likely to sell their stocks at a later opportunity [31]. In consumer purchase decisions, customers showing inertia refrain from making new purchase [32]. In travel behavior, travelers are more likely to stick to their habitual mode, routes and departure time [33-35]. The inertia is also observed in the studies of disease spread and prevention. The influence of inertia is even considered in some SIR models [36]. In a pandemic study focused on the Spanish Flu, the cognitive inertia has been shown as a main reason for the lack of preparation among unaffected areas (Dicke, 2015). The study suggested that cognitive inertia grew from the traditional view that seasonal flu would run for a brief period, and has no threat on healthy group. A recent paper by an infectious disease research team showed the existence of inherent inertia toward social distancing through a game theory model, and stated the important role of national public statements in overcoming the social distancing inertia [37]. Our study provided evidence from observed data on this phenomenon. Answering why such inertia exists is outside the scope of this paper. A future paper can dig deeper and study device-level results to further investigate inertia and learn more about possible causes. Such analysis needs to conform with privacy rules and regulations.

**ACKNOWLEDGMENT**

We would like to thank and acknowledge our partners and data sources in this effort: (1) Amazon Web Service and its Senior Solutions Architect, Jianjun Xu, for providing cloud computing and technical support; (2) computational algorithms developed and validated in a previous USDOT Federal Highway Administration's Exploratory Advanced Research Program project; (3) mobile device location data provider partners; and (4) partial financial support from the U.S. Department of Transportation's Bureau of Transportation Statistics.